\documentclass[prd,twocolumn,superscriptaddress,floatfix,amsmath,amssymb,amsfonts,longbibliography,nofootinbib]{revtex4-2}
\usepackage{amsmath,amssymb,color,graphicx,bm,hyperref,mathrsfs}
\usepackage{physics}
\usepackage{verbatim}

\definecolor{red}{rgb}{0.8,0,0}
\definecolor{RED}{rgb}{0.8,0,0}
\definecolor{violet}{rgb}{0.4,0,0.4}
\definecolor{green}{rgb}{0,0.5,0.0}
\definecolor{GREEN}{rgb}{0,0.5,0.0}
\definecolor{navy}{rgb}{0.0,0.0,0.6}
\definecolor{orange}{rgb}{0.8,0.2,0.0}
\definecolor{blue}{rgb}{0.3,0.0,0.8}

\begin{document}
\title{\textbf{General relativistic heat flow from first order hydrodynamics}}
\author{Bhera Ram}
\email{bhera.ram@iitg.ac.in}
\author{Bibhas Ranjan Majhi}
\email{bibhas.majhi@iitg.ac.in}
\affiliation{Department of Physics, Indian Institute of Technology Guwahati, Guwahati 781039, Assam, India.}

\begin{abstract}
Following the recently proposed stable and causal first-order relativistic hydrodynamics by Bemfica, Disconzi, and Noronha, we find the heat flow equation in the presence of gravity for a non-viscous fluid, which suffers heat dissipation. The derivation is confined to static and stationary backgrounds. We find that in the presence of gravity, the heat flux times a redshift factor is conserved. Then for radial heat flow, the temperature profiles are obtained from the heat equation when the gravity is sourced by Schwarzschild, Schwarzschild-dS, Kerr and Kerr-dS black holes, respectively. Consequently, the chemical potential profile is also discussed.

\end{abstract}
\maketitle	
\section{Introduction}

It is well-established that Zeroth's law gets modified in the presence of gravity. The constancy of temperature across spacetime is modified to $\sqrt{-g_{00}}\, T(x)$ in thermal equilibrium, which is famously known as the Tolman-Ehrenfest (TE) relation \cite{Tolman:1930ona,Tolman:1930zza}. Here, $T(x)$ denotes the temperature of the fluid as measured locally, while $g_{00}$ represents the time-time component of the static metric. Although Tolman and Ehrenfest derived this for static spacetime, it was later demonstrated to apply for stationary backgrounds as well \cite{PhysRev.76.427.2}, when the four velocity of fluid is chosen along a specific timelike Killing vector. However, the stationary situations have also been discussed for a generic choice of timelike Killing vector, leading to modification in TE relation (see e.g. \cite{Santiago:2018lcy}). Following the TE relation, Klein demonstrated that ${\mu}/{T}$ remains constant throughout the spacetime \cite{RevModPhys.21.531}. Here $\mu$ is the local chemical potential of the fluid.  These interesting results then led to several approaches to derive them. Using maximum entropy condition, Cocke \cite{Cocke1965} and later followed by several others \cite{PhysRevD.12.956,Sorkin:1981wd,Gao:2011hh,PhysRevD.85.027503} found those aforesaid relations. A further implication was provided by using Noether symmetry formalism \cite{Green:2013ica,Shi:2021dvd,Shi:2022jya}. Use of an appropriate thermodynamic ensemble has also been done to obtain these results \cite{Roupas:2013fct,Roupas:2014sda,Roupas:2018abp,Roupas:2014hqa}. Rovelli and Smerlek \cite{Rovelli:2010mv} introduced a concept of ``thermal time" to obtain TE relation for a stationary background. For generalization of TE relation at the quantum level, see \cite{Gim:2015era}. 

Recently, using the conservation of energy-momentum, entropy and particle number (for an ideal fluid), Lima \textit{et al.} \cite{Lima:2019brf,Lima:2021ccv} argued that in a static spacetime, when $\mu \neq 0$, TE relation and Klein's law are not satisfied separately. However, using the more viable first-order formalism by Bemfica \textit{et al.}, we in our recent work obtained the definition of local temperature and chemical potential for thermal equilibrium in terms of the acceleration of the fluid. Moreover, it has been observed that further imposition of the local entropy conservation leads to the standard TE and Klein's relation \cite{ram2024}. Also, see \cite{Xia:2023idh} where the TE has been obtained by using equation of state of the fluid.

Usually, if there is a temperature gradient from one spacetime point to another, there will be heat flux. Correspondingly the temperature profile satisfies the heat flow equation. However, in the presence of gravity, it is the differential of TE relation that dictates the flow of heat from one spacetime point to another. In such a case, we expect an analogous heat equation, governing the flow of heat across two spacetime points. In this particular work, we study the governing heat equation in a gravitational background for which we use the well-established first-order causal and stable formalism of relativistic hydrodynamics. Very recently a few works \cite{Valerio2024,Cui:2024fir} in this direction appeared, however, we will discuss later that our analysis is different and also the results are more general.

The formulation of a stable and causal theory describing relativistic viscous fluid has been a long quest. The first two important works were done by Landau \cite{Landau} and Eckart \cite{Eckart:1940te} by adding first-order dissipative contribution to the ideal fluid equations. Although they both describe the Navier-Stokes equation in the non-relativistic limit, they suffer from very fundamental issues of being acausal and unstable about an equilibrium state \cite{Hiscock:1983zz}. Later, Israel and Stewart \cite{Israel:1976tn, Israel:1979wp} added second-order contributions to formulate a stable and casual theory for relativistic hydrodynamics but introduced more degrees of freedom other than the standard ones. In recent years, a concrete first-order theory for relativistic fluid dynamics was proposed by F.S. Bemfica, M. M. Disconzi, J. Noronha \cite{PhysRevX.12.021044} (BDN) based on their earlier works \cite{PhysRevD.98.104064,PhysRevD.100.104020}. After their initiation, Kovtun also dealt with the similar issue in \cite{Kovtun:2019hdm,Hoult:2020eho}.

In this work using BDN formalism (particularly the one, formalized in \cite{PhysRevX.12.021044}), we obtain the heat flow equation in the presence of a gravitational background. The analysis is confined to a non-viscous fluid where the dissipation is governed by heat flux. We find for static and stationary backgrounds that the heat flux current $(q^a)$ times a redshift factor is conserved. Later solving the heat equation for Schwarzschild, Schwarzschild-de Sitter (SdS), Kerr and Kerr-de Sitter (KdS) backgrounds the temperature profiles are obtained in each case. For simplicity, the analysis is done by assuming radial heat flow. It is found that when the heat flux is zero, all the relations reduce to the well-known TE relation. Moreover, we also find the profile for the chemical potential. We feel that the solutions can be used as benchmarks for testing numerical codes
in general relativistic dissipative hydrodynamics.

Before going into the main discussion, let us point out the signature of the background metric and the chosen units, which will be adopted to perform the calculations. Here, we are using $(-,+,+,+)$ as our metric signature and the units are chosen such that the velocity of light in free space is given by $c=1$.

\section{Pre-requisites}
In this work, we follow the choice of variables, taken by BDN. Below, we provide a brief overview of the BDN formalism, as outlined in the original work \cite{PhysRevX.12.021044}. In this formalism, the fluid description is presented as follows.
The constitutive relations that give the baryon current and the energy-momentum tensor are 
\begin{eqnarray}
   &&J^a=n u^a~, 
   \label{eq:A1}
\\   
  &&T^{a b}=(\varepsilon+\mathcal{A}) u^a u^b+(p+\Pi) \Delta^{a b}-2 \eta \sigma^{a b}
  \nonumber
  \\
&&\hspace{1cm}+u^a {q}^b+u^b{q}^a~,
   \label{eq:A2}
\end{eqnarray}
where the expressions for $\mathcal{A}$, $\Pi$, $q^a$ and $\sigma^{a b}$ are as follows:
\begin{eqnarray}
&&\mathcal{A}=\tau_{\varepsilon}\left[u^a \nabla_a \varepsilon+(\varepsilon+p) \nabla_a u^a\right]~,
\label{eq:A3}
\\
&&\Pi=-\zeta \nabla_a u^a+\tau_p\left[u^a \nabla_a \varepsilon+(\varepsilon+p) \nabla_a u^a\right]~,
\label{eq:A4}
\\
&&{q}_b = \frac{  \sigma T(\varepsilon+p)}{n} \Delta_b ^a \nabla_a(\frac{\mu}{T})
\nonumber
\\
&&+\tau_q\left[(\varepsilon+p) u^a \nabla_a u_b+\Delta_b ^a \nabla_a p\right]~,
\label{eq:A5}
\\
&&{\sigma}^{a b}=\frac{1}{2}\Delta^{a c}\Delta^{b d}\left( \nabla_c u_d + \nabla_d u_c - \frac{2}{3}\Delta_{c d}\Delta^{e f}\nabla_e u_f \right)~,
\nonumber
\\
&&=\frac{1}{2}\left(\nabla^a u^b + \nabla^b u^a + u^a \dot{u}^b +  u^b \dot{u}^a  - \frac{2}{3}\Delta^{a b}\nabla_e u^e\right).
\label{eq:A6}
\end{eqnarray}

In the above expressions $\varepsilon$, $s$ $p$, $n$, $T$ and $\mu$ are equilibrium thermodynamic variables -- energy density, entropy, pressure, number density, temperature and chemical potential, respectively. These are connected via the Euler relation $\varepsilon + p = T s + \mu n$ and are consistent with first law of thermodynamics. Further, $u^a$ is a normalized time-like vector (\textit{i.e.} $u_a u^a = -1$), called as the flow or fluid velocity, and $\Delta_{a b} = g_{a b} + u_a u_b$ is a projector onto the space orthogonal to $u^a$.
Additionally, $\tau_\varepsilon$, $\tau_p$, and $\tau_q$ represent the corrections to the out-of-equilibrium components of the energy-momentum tensor, corresponding to the energy density correction $\mathcal{A}$, the bulk viscous pressure $\Pi$, and the heat flux $q^a$, respectively. In the above expressions $\dot{u}^a$ is given by, $\dot{u}^a=u^c\nabla_c u^a$. Furthermore, $\sigma_{ab}$ denotes the traceless shear tensor, with $\zeta$, $\sigma$, and $\eta$ being the coefficients of bulk viscosity, heat conductivity and shear viscosity.

The above description is consistent with the required properties mentioned in \cite{PhysRevX.12.021044}. Particularly the local version of the entropy increase theorem is well satisfied upon using the on-shell conditions $\nabla_a J^a = 0$ and $\nabla_a T^{ab}=0$. The first one gives $u^a\nabla_a n + n\nabla_a u^a = 0$
while the projection of other condition along $u^a$ gives
\begin{eqnarray}
&&u^a\nabla_a \varepsilon + (\varepsilon+p)\nabla_a u^a = - u^a\nabla_a\mathcal{A} - (\mathcal{A}+ \Pi)\nabla_a u^a 
\nonumber
\\
&& - \nabla_a q^a - q^a u^b\nabla_b u_a + 2\eta \sigma_{a b}\sigma^{a b}~,
\label{eq:A9}
\end{eqnarray}
and its projection on a space perpendicular to $u^a$ provides
\begin{eqnarray}
&&(\varepsilon + p)u^a\nabla_a u^b + \Delta^{a b}\nabla_a p = 
-(\mathcal{A} + \Pi)u^a\nabla_a u^b - \Delta^{a b}\nabla_a \Pi 
\notag 
\\
&& + 2\eta \Delta_c^b(\nabla_a \sigma^{a c}) - u^a\nabla_a q^b  - (\nabla_a u^a) q^b - (\nabla_a u^b ) q^a 
\nonumber\\
&&\quad + u^b q^c u^a\nabla_a u_c~.
\label{eq:A10}
\end{eqnarray}
With this short introduction, let us discuss the heat flow equation in the regime of first-order relativistic hydrodynamics using BDN choice of variables.

\section{Equations governing non-viscous heat flow}
Now let us come to discuss the equations governing heat flow in the presence of a gravitational background. In principle the system of equations (\ref{eq:A9}) and (\ref{eq:A10}) along with the baryon current conservation $\nabla_aJ^a=0$ describes the heat convection.
 However, to invoke a specific situation, we consider a non-viscous (\textit{i.e. }absence of shear and bulk viscosity) but non-ideal fluid, where dissipation occurs only due to heat flux. Thus, in the absence of shear and bulk viscosity, the terms associated with the coefficients $\eta$ and $\zeta$ do not arise. However, thermal equilibrium is not maintained -- for instance, in the case of a static background, $\sqrt{-g_{00}}T(x)$ is not constant across different spacetime points -- resulting in heat flow within the fluid.
Under these circumstances, Eq. (\ref{eq:A9}) and Eq. (\ref{eq:A10}) reduce to 
\begin{eqnarray}
&&u^a\nabla_a \varepsilon + (\varepsilon+p)\nabla_a u^a 
\nonumber
\\
&&= - u^a\nabla_a \Big[\tau_{\varepsilon}\big\{u^b \nabla_b \varepsilon+(\varepsilon+p) \nabla_b u^b\big\}\Big]
\nonumber
\\
&& - \Big[(\tau_\varepsilon+\tau_p)\big\{u^b \nabla_b \varepsilon+(\varepsilon+p) \nabla_b u^b \big\}\Big]\nabla_a u^a 
\nonumber
\\
&&- \nabla_a q^a - q^a \dot{u}_a~,
\label{M1}
\end{eqnarray}
and
\begin{eqnarray}
&&(\varepsilon + p)u^a\nabla_a u^b + \Delta^{a b}\nabla_a p 
\nonumber
\\
&&= - (\tau_\varepsilon+\tau_p)\Big[u^a \nabla_a \varepsilon+(\varepsilon+p)\nabla_a u^a\Big]\dot{u}^b
\nonumber
\\
&&-\Delta^{bc}\nabla_c\Big[\tau_p\big\{u^a \nabla_a \varepsilon+(\varepsilon+p)
\nabla_a u^a\big\}\Big] - \nabla_au^a q^b
\nonumber
\\
&&- (\nabla_a u^a)q^b -q^a\nabla_a u^b +  u^b q^a \dot{u}_a~,
\label{M2} 
\end{eqnarray}
for a non-viscous but out-of-thermal equilibrium fluid.

Here, Eq. (\ref{M1}) is of particular interest as it describes heat flow in a non-viscous fluid when it is not in thermal equilibrium. Specifically, for static and stationary backgrounds, substituting Eq. (\ref{eq:A5}) into Eq. (\ref{M1}) yields an equation governing the fluid's temperature and chemical potential gradient as will be shown in the later sections explicitly. This will be more transparent by considering a non-viscous fluid on a static background. Later we will discuss for a stationary background as well.

\subsection{Static background}
We choose the four-velocity of fluid as $u_a = -N\nabla_a t$, where $t$ is the time coordinate and the hypersurface is one of the $t=$ constant surfaces. $N$ can be determined by using the normalization condition $u_au^a=-1$. Here $u^a$ is normal to the hypersurface, and so $\Delta_{ab} = g_{ab}+u_au_b$ is the induced metric. Under these circumstances one can show $\dot{u}_a = \Delta_a^b\nabla_b(\ln\sqrt{N^2}) = \mathscr{D}_a(\ln\sqrt{N^2})$. First, we consider a static background metric 
\begin{equation}
ds^2 = g_{00}dt^2 + g_{\mu\nu} dx^\mu dx^\nu~. 
\label{B3}
\end{equation}
In this case, the metric coefficient $g_{ab}$ are time ($t$) independent and there is no cross term with time differential; \textit{i.e.} $g_{0\mu} = 0$. Also, we have $g^{00} = 1/g_{00}$. Therefore the components of $u^a$ are $u_a = (-N,0,0,0)$ and $u^{a} = (-N g^{00},0,0,0)$ where $N^2 = -g_{00}$. On this static background, the fluid parameters must respect the symmetries of the static background. 
Therefore in coordinates adapted to the timelike symmetry, the fluid scalar parameters do not change along the flow line; \textit{i.e.} $u^a \nabla_a X = 0$, where $X$ stands for any fluid scalar parameter. Therefore, we have $u^a\nabla_a \varepsilon = 0$ along with $\nabla_a u^a = \frac{1}{\sqrt{-g}}\partial_a\left(\sqrt{-g}u^a\right)=0$, as the metric components are time-independent for a static background.
In this situation, Eq. (\ref{M1}) reduces to 
\begin{equation}
\nabla_a q^a + q^a \dot{u}_a = 0~.
\label{Q3}
\end{equation}
Furthermore, $\dot{u}_a$ for the static background is given by 
\begin{eqnarray}
\dot{u}_a &=& \Delta_a^b\nabla_b\left(\ln\sqrt{N^2}\right)  =  (\delta^b_a + u^bu_a)\nabla_b\left(\ln\sqrt{N^2}\right) 
\nonumber
\\
&=& \nabla_a\left(\ln{\sqrt{N^2}}\right)~.
\label{dotu}
\end{eqnarray}
Then (\ref{Q3}) reduces to
\begin{equation}
\nabla_a\left(\sqrt{N^2}\,q^a\right) = 0~.
\label{eq:A18}
\end{equation}
This implies that in the presence of gravity $\sqrt{N^2}q^a$, rather than $q^a$ itself, is the conserved quantity. However, (\ref{eq:A18}) further simplifies to 
\begin{equation}
\partial_a \Big(\sqrt{-g}\sqrt{N^2}~ q^a\Big) = 0~.
\label{B4}
\end{equation}
Next for the static case, it can be shown that the coefficient of $\tau_q$ (\textit{i.e.} the terms within the third bracket) in Eq. (\ref{eq:A5}) can be set to zero, as explicitly shown in Appendix \ref{App2}, resulting in
\begin{equation}
q^a = \alpha \nabla^a\Big(\frac{\mu}{T}\Big)~,
\label{BRM4}
\end{equation}
where $\alpha = \sigma T({\varepsilon+p})/{n}$. In this regard, it should be stressed that we are not setting $\tau_q =0$ in our analysis. Rather we proved that the terms that are appearing with $\tau_q$ in (\ref{eq:A5}) vanish for the present background geometry and the choice of velocity of the fluid. In this sense, our choice of velocity yields BDN theory reminiscent to Eckart's formalism. 
Furthermore, Eq. (\ref{BRM4}) can also be expressed as (see Appendix \ref{AppC} for detailed analysis)
\begin{equation}
q^b = - \kappa\Delta^{a b}\Big(\nabla_a T + T\dot{u}_a\Big)~,
\label{Q2}
\end{equation}
where $\kappa={\sigma(\varepsilon+p)^2}/{(n^2 T)}$.
Use of (\ref{dotu}) in the latter term, yields
\begin{equation}
q_a = -\kappa T \nabla_a\ln{\left(T\sqrt{N^2}\right)}~.
\label{Q5}
\end{equation}
Further, substitution of the above equation into Eq. (\ref{eq:A18}) yields
\begin{equation}
\nabla_a\left[\kappa\,\nabla^a\left(T\sqrt{N^2}\right)\right] = 0~.
\label{eq:A21}    
\end{equation}
The above one can be interpreted as the heat flow equation for a non-viscous fluid under a static gravitational field.
In the case when $\kappa$ is constant ( the same assumption has been considered in \cite{Valerio2024} as well), the above equation reduces to a very nice form $\nabla_a\nabla^a\left(T\sqrt{N^2}\right)= 0$, which further reduces to

\begin{equation}
   \partial_\mu\Big[\sqrt{-g} g^{\mu\nu}\partial_\nu\left(T\sqrt{N^2}\right)\Big]= 0~,
\label{eq:A22}    
\end{equation}
where $\mu$ and $\nu$ stand for only space indices.

For simplicity, we consider a static spherically symmetric metric:
\begin{equation}
ds^2= g_{00} dt^2 + g_{rr} dr^2 + r^2 d\Omega_{(2)}^2~,
\label{B6}    
\end{equation}
as an example of a static metric,
where, ${d}\Omega_{(2)}^2 = d\theta^2 + \sin^2{\theta}d\phi^2 $.
Also, assume that the heat flux is just in the radial direction, \textit{i.e.} $q^a = (0,q^r,0,0)$. Then  Eq. (\ref{B4}) with the use of $\sqrt{-g} = r^2 \sin{\theta}\sqrt{-g_{00} \,g_{rr}}$ for the above metric reduces to 
\begin{equation}
\partial_r \Big( r^2 \sin{\theta}\sqrt{-g_{00} \,g_{rr}}\sqrt{N^2}~ q^r\Big) = 0~. 
\end{equation}
Since the differentiation is with respect to the radial coordinate only, therefore $\sin{\theta}$ can be pulled out leading to
\begin{equation}
\frac{\partial}{\partial r} \Big( r^2\sqrt{-g_{00} \,g_{rr}}\sqrt{N^2}~ q^r\Big) = 0~.
\label{R8}
\end{equation}
Integrating the above along with $N^2 = - g_{00}$ one finds
\begin{equation}
q^r = - \frac{q_0}{r^2 g_{00} \sqrt{g_{rr}}}~,
\label{eq:A27}     
\end{equation}
where $q_0$ is the integration constant. On the other hand, Eq. (\ref{Q5}) yields 
\begin{equation}
q^r = \frac{-\kappa}{g_{rr}\sqrt{-g_{00}}}\,\partial_r\left(T\sqrt{-g_{00}}\right)~.
\label{eq:A28}     
\end{equation}
Next comparing (\ref{eq:A27}) and (\ref{eq:A28}) we find
\begin{equation}
\frac{d}{d r}\left(T\sqrt{-g_{00}}\right) + \frac{q_0\sqrt{g_{rr}}}{\kappa\,r^2 \sqrt{-g_{00}}} = 0~.
\label{eq:A29}     
\end{equation}
One could obtain the same equation by solving Eq. (\ref{eq:A21}) as depicted below. On rewriting Eq. (\ref{eq:A21}), one gets $\partial_a\left[\sqrt{-g}\,\kappa\,g^{ab}\partial_b\left(T\sqrt{N^2}\right)\right] = 0$. Since the heat flow is only radial, thus we only have the radial variation of the temperature profile, therefore for the static background given by Eq. (\ref{B6}), we have $\partial_r\left[\sqrt{-g}\,\kappa\,g^{rr}\partial_r\left(T\sqrt{N^2}\right)\right] = 0 $, which can be further simplified by using the expression for $\sqrt{-g}$ and noting that $g^{rr} = 1/g_{rr}$. Additionally, factoring out the $\sin{\theta}$ as the differentiation is only with respect to $r$ coordinate, we obtain 
\begin{equation}
\frac{d}{d r}\left[\frac{\kappa r^2 \sqrt{-g_{00}}}{\sqrt{g_{rr}}}\frac{d}{d r}\left(T \sqrt{N^2}\right)\right] = 0~.
\label{New1}
\end{equation}
The above equation along with the expression for $N^2$, when integrated once, yields Eq. (\ref{eq:A29}).

Furthermore, Eq. (\ref{eq:A29}) determines the temperature gradient in the absence of thermal equilibrium for a static spherically symmetric background. One can note that when $q_0=0$, we have our usual Tolman relation for a static background. The solution of Eq. (\ref{eq:A29}) gives the profile of temperature for the radial heat flow.

Similarly, use of Eq. (\ref{BRM4}) involving the gradient of $\mu/T$ for radial heat flux, along with the use of (\ref{eq:A27}) yields
\begin{equation}
\frac{\mathrm{d}}{\mathrm{d}r}\Big(\frac{\mu\sqrt{-g_{00}}}{T\sqrt{-g_{00}}}\Big) + \frac{q_0 \sqrt{g_{rr}}}{\alpha r^2 g_{00}} = 0~. 
\label{BRM5}    
\end{equation}
In the above, solution of Eq. (\ref{eq:A29}) can be used to find the radial variation of $\mu\sqrt{-g_{00}}$. This yields the expression for the chemical potential for radial heat flow. In fact use of (\ref{eq:A29}) in (\ref{BRM5}) yields the radial variation equation for $\mu\sqrt{-g_{00}}$ as
\begin{eqnarray}
&&\frac{\mathrm{d}}{\mathrm{d}r}\Big(\mu\sqrt{-g_{00}}\Big) = - \dfrac{q_0\sqrt{g_{rr}}}{r^2\sqrt{-g_{00}}} \Big[\frac{\mu}{T}\frac{1}{\kappa} - \frac{T}{\alpha}\Big]
\nonumber
\\
&& = -\Bigg( \dfrac{q_0^2\,g_{rr}(r)}{r^2}\Bigg)
\nonumber
\\
&&\times \Bigg[\frac{1}{\kappa}\int^{r}\frac{\left(g_{rr}(r')\right)^{3/2}}{\alpha\,r'^2}\,\mathrm{d}r' + \frac{\sqrt{g_{rr}(r)}}{\alpha}\int^{r}\frac{g_{rr}(r')}{\kappa\,r'^2}\,\mathrm{d}r'\Bigg]~.
\label{Klein}
\end{eqnarray}
In the last step (\ref{BRM5}) has been used.
Although the above equation gives us the radial variation for the quantity $\mu\sqrt{-g_{00}}$, it is easier to obtain the expression for $\mu(r)$ by integrating directly Eq. (\ref{BRM5}):
\begin{equation}
\frac{\mu}{T} = A - {q_0}\int\left(\frac{\sqrt{g_{rr}}}{\alpha\,r^2\,g_{00}}\right)\mathrm{d}r~,
\label{Klein2}
\end{equation}
where $A$ is an integration constant. In Appendix \ref{App1} the above equation has been explicitly solved for a few cases assuming $\alpha$ to be constant.

\subsubsection{Application to static backgrounds}
To get a feel of the above results we apply them on a few specific static metrics.

{\it Schwarzschild Geometry: --}
In this section, we take the Schwarzschild geometry as an example of the static spherically symmetric case. On solving Eq. (\ref{eq:A29}) for the Schwarzschild case with $g_{00}= - 1 + \dfrac{2m}{r}$ and $g_{rr} = \dfrac{-1}{g_{00}}$, we get
\begin{equation} 
T(r) = \dfrac{1}{\sqrt{1-\dfrac{2m}{r}}}\Bigg[T_0 - \dfrac{q_0}{2m\kappa}\ln{\left(1-\dfrac{2m}{r}\right)}\Bigg]~,
\label{eq:A32}     
\end{equation}
where $T_0$ is a positive constant. In the above $\kappa$ has been assumed to be a constant. The integration constant $T_0$ can be fixed by demanding, under $r \to$ $\infty$, one should get back the Hawking temperature $\left(T_H = 1/(8\pi m)\right)$ \cite{Hawking:1975vcx} as seen by a static observer, provided the fluid temperature is controlled by that of the horizon.

{\it{Schwarzschild-de Sitter Geometry}: --}
Let us work with Schwarzschild-de Sitter (SdS) spacetime which is a solution of the vacuum Einstein equation with cosmological constant $\Lambda$ and can have two horizons based on the values of the parameters $\Lambda$ and $m$. The geometry for the Schwarzschild-de Sitter is given by the following metric
\begin{equation}
ds^2 = -f(r)dt^2 + \frac{dr^2}{f(r)} + r^2d\Omega_{(2)}^2~,
\label{BR1}
\end{equation}
where $f(r) = 1-\frac{2m}{r}-\frac{\Lambda r^2}{3}$.
For Schwarzschild-de Sitter geometry, we have, $g_{rr} = {-1}/{g_{00}}$ and thus
(\ref{eq:A29}) becomes
\begin{equation}
\frac{\mathrm{d}}{\mathrm{d}r}\left(T\sqrt{-g_{00}}\right) - \frac{q_0}{\kappa\,r^2\,{g_{00}}} = 0~.
\label{eq:A30}     
\end{equation}
Integrating the above equation assuming $\kappa$ to be a constant, gives
\begin{equation}
T\sqrt{-g_{00}} = T_0 - \dfrac{q_0}{\kappa} \int\dfrac{1}{r^2\left(1-2m/r-\Lambda r^2/3\right)}\mathrm{d}r~.
\label{BR3}
\end{equation}
The above equation integrates to
\begin{eqnarray}
T(r)\sqrt{1-\dfrac{2m}{r}-\dfrac{\Lambda r^2}{3}} &=& T_0 + \dfrac{q_0}{2m\kappa}\Bigg(\ln\big|x\big| \label{BR4}\\
\nonumber
&& - \sum_{i}\dfrac{\ln\big|{x-x_{i}}\big|(\Lambda\,r_0^2\,x_i^2 - 3)}{3(\Lambda r_0^2\,x_{i}^2-1)}\Bigg)~;
\end{eqnarray}
where $x\equiv r/r_0$, $r_0$ is the bigger positive root of $f(r)=0$, $x_{i}$ are the solutions to the equation below
\begin{equation}
6 m - 3 r_0 x + \Lambda r_0^3 x^3=0~.
\label{SdS1}
\end{equation}
It can be easily seen that the above equation for the de-Sitter case, Eq. (\ref{BR4}), reduces to the Schwarzschild case given by Eq. (\ref{eq:A32}) under the limit $\Lambda \to 0$.

\subsection{Stationary background}
Let us consider a stationary background given by the following metric 
\begin{equation}
ds^2 = g_{00} dt^2 + g_{0\phi} dt d\phi + g_{rr} dr^2 + g_{\theta\theta}d\theta^2 + g_{\phi\phi}d\phi^2~,
\label{B7}
\end{equation}
having both time translation and azimuthal symmetries, meaning $g_{ab}$ is independent of coordinates $t$ and $\phi$.
Therefore with the choice of $u_a = -N\nabla_a t$, we have $u_a = \left(-N, 0,0,0\right)$. Further, for the above metric, $u^a$ is given by $u^a=\left(-N\,g^{00},0,0,-N\,g^{\phi\,0}\right)$. Demanding $u_a\,u^a = -1$, we have $N^2 ={-1}/{g^{00}}$.

Now since the metric has time translational and azimuthal symmetries, here again we have $\dot{u}_a = \nabla_a\left(\ln{\sqrt{N^2}}\right)$. Moreover, the fluid also respects these symmetries and hence $u^a\nabla_a\varepsilon=0$ along with $\nabla_a u^a = \frac{1}{\sqrt{-g}}\partial_a\left(\sqrt{-g}u^a\right)=0$, as the metric components are $t$ and $\phi$ independent. Therefore, Eq. (\ref{M1}) again reduces to Eq. (\ref{Q3}) for the stationary background as well. Substituting  the expression for $\dot{u}_a$ into Eq. (\ref{Q3}) yields:
\begin{equation}
\nabla_a\Big(\sqrt{N^2} q^a\Big) = \partial_A\Big(\sqrt{-g}\sqrt{N^2} q^A\Big)=0~.
\label{B9}    
\end{equation}
In the above, $\sqrt{-g} = \left\{-g_{rr}g_{\theta\theta} \big(g_{00}g_{\phi\phi} - g_{0 \phi}^2\big)\right\}^{1/2}$ and $A$ stands for radial coordinate $r$ and polar coordinate $\theta$. Therefore here also $\sqrt{N^2} q^a$ is a conserved quantity.

Further, like the static case, we will have exactly the same heat equation, given by Eq. (\ref{eq:A21}). This is because $q^a$ is again given by Eq. (\ref{BRM4}) or Eq. (\ref{Q5}) for the stationary case as well (see Appendix \ref{App2} and \ref{AppC} for detailed discussion). Furthermore, solving Eq. (\ref{B9}) along with Eq. (\ref{Q2}) will provide us with the gradient of the TE relation, which we have explicitly worked out for Kerr and Kerr-de Sitter geometry below.

\subsubsection{Application to stationary metrics}
{\it Kerr Geometry: --}
Let us now turn to an example of a stationary Kerr metric. The metric elements in this case in Boyer-Lindquist coordinates are given by
\begin{eqnarray}
&&g_{00} = - \left(1- \frac{2mr}{\rho^2}\right);\,\,\,\, g_{0\phi} = -\frac{2mar\sin^2\theta}{\rho^2};\,\,\,g_{rr} = \frac{\rho^2}{\Delta}; 
\notag
\\
&&\quad g_{\theta\theta} = \rho^2;\,\,\,\,g_{\phi\phi} = \Big(r^2+a^2+\frac{2ma^2r\sin^2\theta}{\rho^2}\Big)\sin^2\theta,
\label{Kerr}    
\end{eqnarray}
with $\rho^2 \equiv r^2+a^2\cos^2\theta$, $\Delta \equiv r^2-2mr+a^2$, and $a \equiv \frac{J}{m}$~,
where $a$ is the Kerr parameter.
On using Eq. (\ref{Kerr}), we obtain $\sqrt{-g} = \rho^2 \sin{\theta}$.
Again, assuming only radial heat flow, the solution to Eq. (\ref{B9}) is given by
\begin{equation}
q^r = \frac{q_0}{\rho^2\,\sqrt{N^2}}~,
\label{B10}    
\end{equation}
where $N^2 = \dfrac{-1}{g^{00}}$, which for the Kerr metric is given by
\begin{equation}
g^{00}=-\frac{2 a^2 m r \sin ^2(\theta )+\rho ^2 \left(a^2+r^2\right)}{2 a^2 m r \sin ^2(\theta )-\left(a^2+r^2\right) \left(2 m r-\rho ^2\right)}~.
\label{D2}
\end{equation}
Further, on using Eq. (\ref{Q5}), we find
\begin{equation}
\frac{\mathrm{d}}{\mathrm{d}r}\Big(T\sqrt{N^2}\Big) + \frac{q_0}{\kappa g^{rr}\rho^2}= 0~,
\label{B11}    
\end{equation}
for a constant value of $\theta$. This is the counterpart of Eq. (\ref{eq:A29}) for the stationary background.

Similarly on integrating Eq. (\ref{B11}) with $g^{rr}= \dfrac{\Delta}{\rho^2}$ gives
\begin{equation}
T\sqrt{N^2} = T_0 - \int\dfrac{q_0}{\kappa\,\rho^2 g^{rr}} \mathrm{d}r~,
\label{D3}
\end{equation}
which with the assumption of $\kappa$ being a constant yields
\begin{equation}
T\sqrt{N^2} = T_0 - \dfrac{q_0}{\kappa} \int\Bigg(\dfrac{1}{r^2 + a^2 - 2mr}\Bigg)\mathrm{d}r~.
\label{D4}
\end{equation}
On solving Eq. (\ref{D4}), we get
\begin{equation}
T\sqrt{N^2} = T_0 - \dfrac{q_0}{2 \kappa}\dfrac{1}{\sqrt{m^2 - a^2}}\ln\Bigg|{\Bigg(\dfrac{\sqrt{m^2 - a^2}-r+m}{\sqrt{m^2 - a^2}+r-m}\Bigg)}\Bigg|~.
\label{D5}
\end{equation}
Also, the above equation Eq. (\ref{D5}) reduces to the Schwarzschild case given by Eq. (\ref{eq:A32}) under the limit $a\to0$.

{\it Kerr-de Sitter Geometry:--}
The solution to Einstein’s equation with cosmological constant describing a black hole with spin $a$ is called the Kerr-de Sitter (KdS) solution. The geometry for the Kerr-de Sitter background in Boyer-Lindquist coordinates is given by the following metric
\begin{eqnarray}
&&d s^2= \left(r^2+a^2 \cos ^2 \theta\right)\left[\frac{d r^2}{\Delta_r}+\frac{d \theta^2}{1+\frac{\Lambda}{3} a^2 \cos ^2 \theta}\right]
\nonumber
\\
&&+\sin ^2 \theta \left(\frac{1+\frac{\Lambda}{3} a^2 \cos ^2 \theta}{r^2+a^2 \cos ^2 \theta}\right)\left[\frac{a d t-\left(r^2+a^2\right) d \phi}{1+\frac{\Lambda}{3} a^2}\right]^2
\nonumber
\\
&& -\frac{\Delta_r}{\left(r^2+a^2 \cos ^2 \theta\right)}\left[\frac{d t-a \sin ^2 \theta d \phi}{1+\frac{\Lambda}{3} a^2}\right]^2~,
\end{eqnarray}
where 
\begin{eqnarray}
&&\rho^2 \equiv r^2 + a^2\cos^2{\theta}~,
\nonumber
\\
&&\Delta_r \equiv \left(r^2+a^2\right)\left(1-\frac{\Lambda}{3} r^2\right)-2 m r~,
\nonumber
\\
&&\hspace{0.5cm}\equiv r^2-2 m r+a^2-\frac{\Lambda r^2}{3}\left(r^2+a^2\right)~.
\label{K0}
\end{eqnarray}
On using the above metric, we have $\sqrt{-g} = {\rho^2 \sin{\theta}}/{\left(1+\frac{\Lambda a^2}{3}\right)^2}$.
Similarly, on solving Eq. (\ref{B9}) for the Kerr-de Sitter metric, assuming radial heat flow gives
\begin{equation}
q^r = \frac{q_0\,\left(1+\frac{\Lambda a^2}{3}\right)^2}{\rho^2\,\sqrt{N^2}}~,
\label{K1}
\end{equation}
where $N^2 = \dfrac{-1}{g^{00}}$, which for the Kerr-de Sitter metric is given by
\begin{multline}
g^{00}=\dfrac{-\Bigg(1+\dfrac{\Lambda a^2}{3}\Bigg)^2}{\Delta_r \rho^2 \Bigg(1+\dfrac{\Lambda a^2 \cos{\theta}^2}{3}\Bigg)}\Bigg[\Big(a^2+r^2\Big)^2\Big(1+\dfrac{\Lambda a^2 \cos^2{\theta}}{3}\Big)\\
-\Delta_r\,a^2\,\sin^2{\theta}\Bigg]~.
\label{K2}     
\end{multline}
Further, on using Eq. (\ref{Q5}), we have
\begin{equation}
\frac{\mathrm{d}}{\mathrm{d}r}\Big(T\sqrt{N^2}\Big) + \frac{q_0\,\left(1+\frac{\Lambda a^2}{3}\right)^2}{\kappa g^{rr}\rho^2}= 0~,
\label{K3}    
\end{equation}
for a constant value of $\theta$.
Integrating Eq. (\ref{K3}) with $g^{rr}= \dfrac{\Delta_r}{\rho^2}$, assuming $\kappa$ being a constant yields
\begin{equation}
T\sqrt{N^2} = T_0 - \dfrac{q_0\,\left(1+\frac{\Lambda a^2}{3}\right)^2}{\kappa}\displaystyle{\int} \Bigg(\dfrac{1}{\Delta_r}\Bigg)\mathrm{d}r~,
\label{E4}
\end{equation}
which reduces to
\begin{multline}
T\sqrt{N^2} = T_0 - \dfrac{q_0\,\left(1+\frac{\Lambda a^2}{3}\right)^2}{\kappa}\\
\int \Bigg(\dfrac{1}{r^2-2 m r+a^2-\frac{\Lambda r^2}{3}\left(r^2+a^2\right)}\Bigg)\mathrm{d}r~.
\label{E5}
\end{multline}
Integrating the above equation yields
\begin{eqnarray}
T\sqrt{N^2} &=& T_0  + \dfrac{q_0\,\left(1+\frac{\Lambda a^2}{3}\right)^2}{\kappa}
\nonumber
\\
&\times&\dfrac{3}{2}\sum_i \Biggl(\dfrac{\log \big|x - x_i\big|}{2 \Lambda r_0^3\,x_i^3 + \Lambda a^2\,r_0\,x_i - 3 r_0\,x_i + 3 m} \Biggr)~,
\nonumber
\\
\label{Int}
\end{eqnarray}
where $x\equiv r/r_0$, $r_0$ is the bigger positive root of $\Delta_r=0$, $x_{i}$ are the solutions to the equation 
\begin{equation}
-3a^2 + {6 r_0}{m}x - 3 { r_0^2}x^2 + \Lambda a^2 r_0^2 x^2 + \Lambda r_0^4 x^4 = 0~.
\label{root}
\end{equation}
Similar to the previous case it reduces to the Schwarzschild-de Sitter and Kerr cases given by the equations Eq. (\ref{BR4}) and Eq. (\ref{D5}) under the limit $a \to 0$ and $\Lambda \to 0$, respectively.

\section{Further consequences}
For a static background (\ref{B4}) implies that the rate of heat accepted or released by a cross-sectional area perpendicular to the heat flux direction of the fluid volume is given by
\begin{equation}
\frac{dQ}{dt} = \pm \int_{\partial \mathcal{V}} \sqrt{-g}\sqrt{N^2} q^\mu d^2\Sigma_\mu~,
\label{B5}    
\end{equation}
where $d^2\Sigma_\mu = d^2x_\perp \hat{n}_\mu$. Here $+$ ($-$) sign signifies heat absorbed (released) by the surface $\partial\mathcal{V}$ and $\hat{n}^\mu$ is the unit outward normal on it. 

The above structure is justified by the following discussion. Since for the static and stationary backgrounds,  Eq. (\ref{B4}) reduces to $\partial_\mu \Big(\sqrt{-g}\sqrt{N^2}~ q^\mu\Big) = 0$ because of the time independence, where $\mu = 1,2,3.$ Therefore, it is valid at each space point, meaning we can integrate over the space volume which is given by a $t=$ constant hypersurface to have a vanishing value. Then we have,
\begin{equation}
\int_{V_t}d^3x \partial_\mu \Big(\sqrt{-g}\sqrt{N^2}~ q^\mu\Big) = 0~,
\label{New2}
\end{equation}
which can be converted to a two-dimensional surface integration by using the Gauss divergence theorem. This is given by 
\begin{equation}
\oint_{\partial V_t} \Big(\sqrt{-g}\sqrt{N^2}~ q^\mu\Big) d\Sigma_\mu   = 0~.
\label{New3}
\end{equation}
The above one consists of three spacelike surfaces which are defined by $r=$ constant, $\theta=$ constant and $\phi=$ constant.
Since we have only radial heat flow, the term that survives is given by 
\begin{equation}
\int_{r=\text{constant}} \Big(\sqrt{-g}\sqrt{N^2}~ q^r\Big) d\Sigma_r   = 0~.
\label{New4}
\end{equation}
Now if the $r=$ constant surface consists of $r=r_1$ (constant) and $r=r_2$ (constant) surfaces which are at finite radial distances (e.g. surfaces of two different spheres with radii $r_1$ and $r_2$) and the space volume $V_t$ is the region in between them, then the above one 
yields
\begin{equation}
\int_{r_1} \Big(\sqrt{-g}\sqrt{N^2}~ q^r\Big) d\Sigma_r   = \int_{r_2} \Big(\sqrt{-g}\sqrt{N^2}~ q^r\Big) d\Sigma_r ~.
\label{New5}
\end{equation}
In the above, each term corresponds to the flux of heat along the radial direction through the cross-section of the respective $r=$ constant surface. This is exactly the term of the right hand side in (\ref{B5}) with $\hat{n}_\mu = \hat{n}_r$.


Let us now concentrate on (\ref{B5}). For the metric (\ref{B6}) with only radial heat flux, it reduces to
\begin{equation}
\frac{dQ}{dt} = \pm \int d\theta d\phi q_0 \sin\theta = \pm 4\pi q_0~,
\label{BRM1}
\end{equation}
on a $r=$ constant cross-section of the hypersurface. In the first equality we have used $\hat{n}_\mu = (0,1,0,0)$ and (\ref{eq:A27}). In the second equality, $q_0$ is assumed to be a pure constant. Now if two spacetime points are at different temperatures, say $T_1$ and $T_2$ with $T_1\sqrt{g_{00}(r_1)}>T_2\sqrt{-g_{00}(r_2)}$, then one with larger TE value will release heat and another one will absorb the same. This situation can arise when spacetime has two horizons, like in Schwarzschild de-Sitter spacetime. Then the change of entropy of the first point is $dS_1/dt =  (1/T_1)(dQ_1/dt) = -(4\pi q_0)/T_1$ and the same for other one is $dS_2/dt = (1/T_2)(dQ_2/dt) = (4\pi q_0)/T_2$. Here $q_0$ has been chosen to be positive. Note that for both locations, the rate of change of heat is numerically the same because of the conservation equation (\ref{B4}). Then the total entropy change is given by
\begin{eqnarray}
\frac{dS}{dt} &=& \frac{dS_1}{dt} + \frac{dS_2}{dt}
\nonumber
\\
&=& 4\pi q_0\Big(\frac{1}{T_2} - \frac{1}{T_1}\Big)~.
\label{BRM2}
\end{eqnarray}
However, if the above one is evaluated with respect to the proper time $d\tau = dt/u^0= dt/\sqrt{-g^{00}} = dt~\sqrt{-g_{00}} $ for the static metric, then we have
\begin{eqnarray}
\Delta_\tau S &=& \frac{1}{\sqrt{-g_{00}(r_1)}}\frac{dS_1}{dt} + \frac{1}{\sqrt{-g_{00}(r_2)}} \frac{dS_2}{dt}
\nonumber
\\
&=& 4\pi q_0\Big(\frac{1}{T_2\sqrt{-g_{00}(r_2)}} - \frac{1}{T_1\sqrt{-g_{00}(r_1)}}\Big)~,
\label{BRM3}
\end{eqnarray}
where $q_0/(T\sqrt{-g_{00}})$ can be determined from (\ref{eq:A29}). Note that the above one is consistent with the second law of thermodynamics as we have $q_0>0$ and  $T_1\sqrt{-g_{00}(r_1)} > T_2\sqrt{-g_{00}(r_2)}$. Interestingly the total entropy change per unit proper time is determined not only through the local temperatures, but also the redshift factors plays an important role.   


We will see that the similar applies to stationary cases as well. Since here also the conserved quantity is $\sqrt{N^2}q^a$, (\ref{B5}) retains. Consequently, use of (\ref{B10}) and the expression for the metric determinant for Kerr, and similarly (\ref{K1}) and the metric determinant for KdS, one finds (\ref{BRM1}) for both the cases. 
However, in this case the proper time is given by $d\tau = dt/u^0 = dt/\sqrt{-g^{00}}$ which is not equal to $dt \sqrt{-g_{00}}$ as $g^{00}\neq 1/g_{00}$ for stationary background. Therefore, we have 
\begin{eqnarray}
\Delta_\tau S &=& {\sqrt{-g^{00}(r_1)}}\frac{dS_1}{dt} + {\sqrt{-g^{00}(r_2)}} \frac{dS_2}{dt}
\nonumber
\\
&=& 4\pi q_0\Big(\frac{\sqrt{-g^{00}(r_2)}}{T_2} - \frac{\sqrt{-g_{00}(r_1)}}{T_1}\Big)~,
\label{N90}
\end{eqnarray}
which is identical to static case when one replaces $g^{00} = 1/g_{00}$ in (\ref{N90}). The expression for $\dfrac{\sqrt{-g^{00}(r)}}{T}$ can be obtained from Eq. (\ref{D5}) and Eq. (\ref{Int}) for Kerr and KdS geometry, respectively.

\section{Conclusions}
Applying the recently proposed causal and stable first-order relativistic hydrodynamics we discussed the phenomena of heat flow in presence of gravity. Particularly, we confined our analysis to static and stationary backgrounds for a specific choice of fluid four-velocity. The fluid has been chosen as non-viscous, in which the heat dissipation is occurring solely due to the gradient of TE relation. We observed that similar to the modification of the conventional Zeroth's law by a redshift factor in the presence of gravity, the conserved quantity due to heat flow is the standard heat flux times the same redshift factor, as described in equation (\ref{eq:A18}). The heat flow equation, describing the temperature profile, has been derived. We explicitly solved it for Schwarzschild, SdS, Kerr and KdS backgrounds providing examples for each case respectively and found the temperature profile under these circumstances. The solutions have been obtained by assuming only radial heat flux. Also the profile for chemical potential has been addressed. Moreover, we then showed that all the expressions reduce to the usual TE relation and Klein's law when the heat flux is absent.

During this work, we came across a similar work \cite{Valerio2024}. However, their formulation is within Eckart's first-order formalism. Moreover in their analysis, they have worked with the equation of state and have simultaneously used both equations related to the conservation of fluid energy-momentum tensor. On the other hand, our analysis is based on much viable first-order relativistic hydrodynamics, given by BDN, which takes care of all the shortcomings of Eckart's formalism and is stable and causal. Furthermore, the calculation does not require the use of any equation of state. Additionally, their analysis is confined to static cases, however, we have included stationary backgrounds as well. Also in \cite{Cui:2024fir} the subject of the heat flow equation has been discussed. However, the calculation was done for a charged relativistic gas in the presence of electromagnetic and gravitational fields based on a different approach. On the other hand, we considered a generic fluid, described by first-order BDN formalism.

Although the present analysis features the aspects of heat flow in presence of gravity, the results, as stated earlier, have restricted applications due to the imposition of various conditions. For example, the four-velocity $u_a$ can be chosen in much general way where it will have both time and space components. However, we have chosen a particular form where the fluid flow is normal to $t=$ constant hypersurface, where $t$ has been taken as the time coordinate of the metric. Thus $u_a$ only has time component for the present analysis. Now this particular flow in the literature is known as `normal flow'. Such choice has been taken in various cases, for instance, see \cite{Santiago:2018lcy}. It will be really interesting to extend our analysis to any general choice of $u_a$, but that analysis has to be done separately. Thus, our results are valid for normal flow of fluid. Furthermore, here we have solved the equations for the radial heat flow, which provided analytic expressions for the profiles of the temperature and chemical potential. Such a solution, although related to a simple case, can provide a realization of the nature of a particular type of heat flow.  This situation can be interesting in physically relevant scenarios, e.g. if two space points differ only by radial distance, then how the heat flow will take place can be obtained from this analysis. Situations like this were also discussed in \cite{Faraoni:2023gqg}. It would be interesting and important for practical purposes to investigate the generalization of these analyses. The particular aim would be to check how much restrictions can be lifted to perform an analytical analysis. Our investigations in these directions are going on and we hope that we will be able to provide more input in the future.

\begin{acknowledgments}
The work of BR is supported by the University Grants Commission (UGC), Government of India, under the scheme Junior Research Fellowship (JRF). BRM is supported by Science and Engineering Research Board (SERB), Department of Science $\&$ Technology (DST), Government of India, under the scheme Core Research Grant (File no. CRG/2020/000616). We also thank the anonymous referee for various important comments which led to an extensive improvement of the manuscript.
\end{acknowledgments}

\appendix
\section{Steps leading to Eq. (\ref{BRM4})} \label{App2}
\subsection{Static background }
As discussed in Section III, the equations governing the conservation of the energy-momentum tensor, along with the conditions for a non-viscous fluid, are given by Eq. (\ref{M1}) and Eq. (\ref{M2}).
For a static background given by Eq. (\ref{B3}), we have $u^a\nabla_a X = 0$ and $\nabla_a u^a = 0$, where $X$ represents any scalar fluid variable. Consequently, under these conditions, Eq. (\ref{M2}) simplify to:
\begin{eqnarray}
(\varepsilon + p)u^a\nabla_a u^b &+& \Delta^{a b}\nabla_a p =
\nonumber
\\
&& - u^a\nabla_a q^b -q^a\nabla_a u^b +  u^b q^a \dot{u}_a~.
\label{N4} 
\end{eqnarray}
Let us now focus on the right-hand side of the above equation. As the vector field $u^a$ is hypersurface orthogonal (since our choice of $u_a$ is $u_a = -N\nabla_at$), it satisfies the Frobenius theorem:
\begin{equation}
u_{[a}\nabla_b u_{c]}=0~.
\label{Frob}
\end{equation}
On contracting Eq. (\ref{Frob}) with $u^a$, we get
\begin{equation}
-\nabla_b u_c + u_c \dot{u}_b - u_b \dot{u}_c + \nabla_c u_b = 0~.
\label{Frob1}
\end{equation}
The above equation can also be interpreted as follows.
Usually, the velocity gradient can be decomposed as
\begin{equation}
\nabla_b u_a=\omega_{a b}+\sigma_{a b}+\frac{1}{3} (\nabla_cu^c) \Delta_{a b}-\dot{u}_a u_b~,
\label{R1}
\end{equation}
where
\begin{equation}
\omega_{a b}  = \frac{1}{2}\left(\nabla_b u_a - \nabla_a u_b + \dot{u}_a u_b - \dot{u}_b u_a \right)~.
\label{R2}
\end{equation}
Further, it is known that for $u_a$ to be hypersurface orthonormal, $w_{ab} = 0$, which again provides us with Eq. (\ref{Frob1}). 

On further contraction of Eq. (\ref{Frob1}) with $q^b$ yields
\begin{equation}
-q^b\nabla_b u_c + u_c\dot{u}^b q_b + q^b\nabla_c u_b = 0~.
\label{Frob2}
\end{equation}

Use of Eq. (\ref{Frob2}) into Eq. (\ref{N4}), gives
\begin{eqnarray}
(\varepsilon + p)u^a\nabla_a u^b + \Delta^{a b}\nabla_a p &=& -g^{bc}\left(u^a\nabla_a q_c + q^a\nabla_c u_a\right)
\nonumber
\\
&=& -g^{bc}\pounds_u q_c~,
\label{Frob3} 
\end{eqnarray}
where $\pounds_{u}$ denotes the Lie variation along $u^a$.
If $\xi^a$ is the timelike Killing vector corresponding to the time translational symmetry for the static metric (\ref{B3}), then it is given by $\xi^a = (1,0,0,0)$. Since $u^{a} = (-N g^{00},0,0,0)$, $u^a$ is proportional to $\xi^a$. Therefore, we have $u^a = \Tilde{N}\xi^a$ with $\Tilde{N}=-N g^{00}$. Thus the right-hand side of the above equation can be written as
\begin{eqnarray}
\pounds_u q_c &=& u^a\nabla_a q_c + q_a\nabla_c u^a~
\nonumber
\\
&=& \Tilde{N}\xi^a\nabla_a q_c + \Tilde{N}q_a\nabla_c \xi^a + \xi^a q_a \nabla_c\Tilde{N}~.
\label{Frob4}
\end{eqnarray}
The last term vanishes because $q^a$ is a spatial vector on the hypersurface perpendicular to $\xi^a$. As a result, the above one simplifies to
\begin{equation}
\pounds_u q_c=\Tilde{N}\Big(\xi^a\nabla_a q_c + q_a\nabla_c \xi^a\Big)  = \Tilde{N}\pounds_\xi q_c~.
\label{Frob6}
\end{equation}  
Now since the fluid quantities must respect the symmetries of the background spacetime, we have $\pounds_\xi q_c =0$. Therefore one finds $\pounds_u q_c=0$ and hence right hand side of Eq. (\ref{Frob3}) vanishes. Thus we find
\begin{eqnarray}
(\varepsilon + p)&u&^a\nabla_a u^b + \Delta^{a b}\nabla_a p =
 0~.
\label{Frob7} 
\end{eqnarray}
Finally use of Eq. (\ref{Frob7}) into Eq. (\ref{eq:A5}), provides us the expression for the heat current $q^a$, as given in (\ref{BRM4}) for the static background. 
The same result is demonstrated for the stationary case in the following analysis.

\subsection{Stationary background}
Again the non-viscous fluid equation, given by Eq. (\ref{M2}) for the stationary background (\ref{B7}), simplifies to Eq. (\ref{N4}) due to $u^a\nabla_a \varepsilon = 0$ and $\nabla_a u^a = 0$. Since in this case our choice of $u_a$ also satisfies the Frobenius theorem, we will have Eq. (\ref{Frob3}) for the stationary background as well. The background has two Killing vectors, $\xi_{(t)}^b = (1,0,0,0)$ and $\xi_{(\phi)}^b = (0,0,0,1)$, corresponding to the time translation and azimuthal symmetries, respectively. Further, $u^b$ is given by $u^b=\left(-N\,g^{00},0,0,-N\,g^{\phi\,0}\right)$, thus we can express $u^b$ as $u^b = N_1\xi_{(t)}^b + N_2\xi_{(\phi)}^b$ with $N_1 = -N g^{00}$ and $N_2 = -N g^{0\phi}$, respectively. Then, we have
\begin{eqnarray}
\pounds_u q_c &=& N_1\pounds_{\xi_{(t)}}q_c + N_2\pounds_{\xi_{(\phi)}}q_c + \big(q_b\xi_{(t)}^b\big)\nabla_cN_1
\nonumber
\\
&&+ \big(q_b\xi_{(\phi)}^b\big)\nabla_cN_2~.
\label{Appn1}
\end{eqnarray}
Furthermore, since we consider only radial heat flow in our analysis, \textit{i.e.} $q^a = (0, q^r, 0, 0)$, the covariant components in the stationary background are $q_a = (0, q_r, 0, 0)$. Therefore, the last two terms vanish since $q_b \xi_{(t)}^b = q_r \xi_{(t)}^r = 0$ due to $\xi_{(t)}^r = 0$, and similarly, $q_b \xi_{(\phi)}^b = q_r \xi_{(\phi)}^r = 0$ as $\xi_{(\phi)}^r = 0$. This leaves only the first two terms in Eq. (\ref{Appn1}). These are individually zero as the fluid quantities must respect the symmetries of the background metric, thereby confirming that $\pounds_u q_c = 0$, resulting in Eq. (\ref{Frob7}) for the stationary case as well. This provides the expression (\ref{BRM4}) for the heat current in the stationary background, which is identical to that of the static case.

\section{Steps leading to Eq. (\ref{Q2})}
\label{AppC}
In this section, we show how $q^b$ can be expressed by Eq. (\ref{Q2}) for static and stationary backgrounds.
The use of Euler's relation $\varepsilon + p = T\,s + \mu\,n$ along with first law of thermodynamics $d\varepsilon = Tds + \mu dn$, yields
\begin{equation}
\nabla_a p = \left(\frac{\varepsilon+p}{T}\right)\nabla_a T +  n T\,\nabla_a\left(\frac{\mu}{T}\right)~.
\label{App4}
\end{equation}
Further, substituting Eq. (\ref{App4}) into Eq. (\ref{M2}) and rearranging the terms slightly, we obtain
\begin{multline}
\sigma\,T\left(\frac{\varepsilon+p}{n}\right)\Delta^{ab}\nabla_a\left(\frac{\mu}{T}\right)\\
= - \frac{\sigma(\varepsilon+p)^2}{n^2\,T}\bigg(\Delta^{a b}\nabla_a T + T\dot{u}^b\bigg)\\
+\frac{\sigma(\varepsilon+p)}{n^2}\bigg[-(\tau_\varepsilon+\tau_p)\Big(u^a \nabla_a \varepsilon+(\varepsilon+p)\nabla_a u^a\Big)\dot{u}^b\\
-\Delta^{bc}\nabla_c\bigg\{\tau_p\Big(u^a \nabla_a \varepsilon+(\varepsilon+p)
\nabla_a u^a\Big)\bigg\}- u^a\nabla_a q^b \\
- (\nabla_a u^a)q^b -q^a\nabla_a u^b +  u^b q^a \dot{u}_a\bigg]~.
\label{App6} 
\end{multline}
By imposing the static and stationary conditions \textit{i.e. }$u^a\nabla_a\varepsilon=0$ and $\nabla_a u^a =0$, Eq. (\ref{App6}) simplifies to
\begin{multline}
\sigma\,T\left(\frac{\varepsilon+p}{n}\right)\Delta^{ab}\nabla_a\left(\frac{\mu}{T}\right)\\
= - \frac{\sigma(\varepsilon+p)^2}{n^2\,T}\bigg(\Delta^{a b}\nabla_a T + T\dot{u}^b\bigg)\\
- u^a\nabla_a q^b - q^a\nabla_a u^b +  u^b q^a \dot{u}_a~,
\label{App7} 
\end{multline}
which on further use of Eq. (\ref{Frob2}), reduces to 
\begin{multline}
\sigma\,T\left(\frac{\varepsilon+p}{n}\right)\Delta^{ab}\nabla_a\left(\frac{\mu}{T}\right)=
-
\kappa\bigg(\Delta^{a b}\nabla_a T + T\dot{u}^b\bigg)\\
-g^{bc}\pounds_u q_c~.
\label{App8}    
\end{multline}
where $\kappa={\sigma(\varepsilon+p)^2}/{(n^2 T)}$. Now, as discussed in Appendix \ref{App2}, for both static and stationary backgrounds, we have $\pounds_u q_c = 0$. Therefore, the above analysis shows the equivalence of Eq. (\ref{BRM4}) and Eq. (\ref{Q2}).

\section{Determining chemical potential}\label{App1}
\subsection{Schwarzschild black hole}
Solving Eq. (\ref{Klein2}) for the Schwarzschild case with $g_{00} = {-1}/{g_{rr}} = -\left(1-\frac{2m}{r}\right)$ and assuming $\alpha$ to be constant, we have
\begin{equation}
\frac{\mu}{T} = A + \frac{q_0}{\alpha}\int\frac{1}{r^2\left(1-\frac{2m}{r}\right)^{3/2}}\mathrm{d}r~.
\label{Klein3}
\end{equation}
Further, on using Eq. (\ref{eq:A32}), we have 
\begin{multline}
\left(\mu(r)\,\sqrt{1-\frac{2m}{r}}\right)_{(sch)} =  \Bigg(T_0 - \frac{q_0}{2 m \kappa}\ln{\left(1 - \frac{2m}{r}\right)}\Bigg)\\
\times \Bigg(A - \frac{q_0}{m\,\alpha}\frac{1}{\sqrt{1- \frac{2m}{r}}}\Bigg)~,
\label{musch}
\end{multline}
where $A$ is the integration constant.

\subsection{Schwarzschild-de Sitter black hole}
Again for the Schwarzschild-de Sitter case, Eq. (\ref{Klein2}) assuming $\alpha$ to be constant reduces to,
\begin{equation}
\frac{\mu}{T} = A + \frac{q_0}{\alpha}\int\frac{1}{r^2\left(1-\frac{2m}{r} - \frac{\Lambda r^2}{3}\right)^{3/2}}\mathrm{d}r~,
\label{Klein4}
\end{equation}
which doesn't have a simplified or compact form. However, it can be solved near the horizon, using $r\to r'+r_b$, where $r_b$ is the black hole horizon.
The above equation near the black hole horizon reduces to
\begin{equation}
\frac{\mu}{T} = A + \frac{q_0}{\alpha}\int\frac{1}{(r'+r_b)^2\,\Big(2\,\kappa_H\,r'\Big)^{3/2}}\mathrm{d}r'~,
\label{Klein5}
\end{equation}
where $f'(r_b) = 2\kappa_H$ with $\kappa_H$ is the surface gravity. On further simplifications, it yields
\begin{equation}
\frac{\mu}{T} = A + \frac{q_0}{\alpha\,(2\kappa_H)^{3/2}}\int\frac{1}{(r'+r_b)^2\,\big(r'\big)^{3/2}}\mathrm{d}r'~,
\label{Klein6}
\end{equation}
which on integrating, gives
\begin{eqnarray}
&&\left(\mu\sqrt{-g_{00}}\right)_{SdS} =\\
\nonumber
&&\left(T\sqrt{-g_{00}}\right)_{SdS}\Bigg[A - \frac{q_0}{\alpha\,(2\kappa_H)^{3/2}}\Bigg(\frac{2r_b + 3 r'}{\sqrt{r'}\left(r_b^3 + r_b^2 r'\right)}\\
\nonumber
&& + \frac{3}{r_b^{5/2}}\arctan\sqrt{\frac{r'}{r_b}}\Bigg)\Bigg]~.
\label{Klein7}
\end{eqnarray}
Note that this is valid near the black hole horizon, where $\left(T\sqrt{-g_{00}}\right)_{SdS}$ is determined by the near horizon expression of Eq. (\ref{BR4}).

Here, we avoid presenting the same for Kerr and Kerr-dS as the expressions are becoming very cumbersome, even within the near-horizon approximation. However, in principle, these can be determined by performing similar integrations.

\bibliographystyle{apsrev}

\bibliography{bibtexfile}
\end{document}